\documentclass[twocolumn,showpacs,showkeys,floatfix]{revtex4}
\usepackage{amssymb}
\usepackage{amsmath}
\usepackage{graphicx}
\usepackage{color}
\usepackage{ulem}

\newcommand{\td}[1]{\frac{{d}{#1}}{{d}t}}

\newcommand{\bs}[1]{\left[{#1}\right]}

\newcommand{\bb}{\begin{eqnarray}}
\newcommand{\ee}{\end{eqnarray}}
\newcommand{\nn}{\nonumber}

\begin{document}

\title{Interaction Effects on the Size Distribution in a Growth Model}

\author{Jungzae \surname{Choi}}
\affiliation{Department of Physics and Department of Chemical Engineering, Keimyung University, Daegu 42601, Korea}
\author{M.Y. \surname{Choi}}
\affiliation{Department of Physics and Astronomy and Center for Theoretical Physics, Seoul National University, Seoul 08826, Korea}
\author{Byung-Gook \surname{Yoon}}
\thanks{E-mail: bgyoon@ulsan.ac.kr}
\affiliation{Department of Physics, University of Ulsan, Ulsan 44610, Korea}

\begin{abstract}
We study, both analytically and numerically, the interaction effects on the skewness of the size distribution of elements in a growth model. We incorporate two types of global interaction into the growth model, and develop analytic expressions for the first few moments from which the skewness of the size distribution is calculated.
It is found that depending on the sign of coupling, interactions may suppress or enhance the size growth, which in turn leads to the decrease or increase of the skewness. The amount of change tends to increase with the coupling strength, rather irrespectively of the details of the model.
\end{abstract}


\maketitle

\section{Introduction}
Many phenomena exhibit asymmetric skew distributions of characteristic spectra, having long tails to one side; among them power-law, log-normal, and Weibull distributions appear most ubiquitously in a variety of systems. The power-law distribution, well-known examples of which include Zipf's law~\cite{Zipf1949} and Pareto's law~\cite{Pareto1896}, manifests scale invariance of the system and observed in  various physical, biological, economical, and social systems~\cite{Newman2005}.
A variety of mechanism such as the Yule process~\cite{Yule1925}, Gibrat's law~\cite{Gibrat1931}, or preferential attachment~\cite{Barabasi199} has been proposed to explain the power-law distribution.
Related to the power-law distribution, the log-normal distribution, often used as an alternative~\cite{Jo2007}, is known to emerge from multiplication of a large number of independent random variables~\cite{Limpert2001}. The Weibull distribution also arises in a variety of systems~\cite{Jo2007,Peterson1985,McDowell1996}, and is attributed to diverse origins, e.g., fractal cracking~\cite{Peterson1985}, extreme value statistics~\cite{Beirlant2004} and Fickian diffusion~\cite{Kosmidis2003}.

Lacking theoretical understanding of the skew distributions that emerge in evolving systems, there has been progress to elucidate the emergence of skew distributions in a general framework of the master equation approach~\cite{Choi2009,Goh2010}. This approach illustrates how log-normal, power-law, and Weibull distributions arise in a system of noninteracting elements, depending on the detailed condition such as growth, production, and division of each element. Further, the entropy of the system has also been obtained via this approach~\cite{Goh2017}.

In a real system, however, interactions between elements are present and may affect the evolution of the size distribution of the system. In this work we consider a growth model for an evolving system of elements interacting with each other, and study the interaction effects on the evolution of the size distribution, focusing especially on the skewness of the distribution.
To probe this issue, we incorporate two types of interaction into the growth model of Refs.~\cite{Choi2009,Goh2010}: one proportional to the size difference between elements and the other proportional to the product of the sizes. It is found that interactions may enhance or suppress the size growth, depending on the sign of the coupling. In the latter case of suppressing the growth, interactions also reduce skewness of the distribution, regardless of the details of the model.

This paper consists of five sections:
In Sec. II, we introduce the growth model system where elements interact with each other. Specifically, two types of global interaction are considered and the corresponding time evolution equations are derived.
Section III is devoted to analytic expressions for the first three moments of the size variable if possible, from which the skewness is calculated for both types of interaction.
Also described is how to obtain the distribution function via numerical integration.
Results are presented and discussed in Sec. IV. Finally, Sec. V gives a brief summary.

\section{Growth Model}

We consider a system of $N$ elements, the $i$th of which is characterized by its size $x_i$ ($i = 1, \ldots, N$).
The configuration of the system is specified by the sizes of all elements, $\{x_1^{}, \dots, x_N^{}\}$.  The probability $P(x_1^{}, \dots, x_N^{}; t)$ for the system to be in configuration $\{x_1^{}, \dots, x_N^{}\}$ at time $t$ is governed by the master equation
\begin{widetext}
  \begin{equation}
    \label{eq: master equation}
      \td{} P(x_1^{},\dots,x_{N}^{};t) = \sum_{i=1}^{N} \int d x_i'\, \bs{\omega(x_i' \rightarrow x_i^{}) P(x_1^{},\dots,x_i',\dots,x_{N}^{};t) - \omega(x_i \rightarrow x_i') P(x_1^{},\dots,x_{N}^{};t)} ,
  \end{equation}
\end{widetext}
where $\omega(x_i \rightarrow x_i')$ is the transition rate for the $i$th element to change its size from $x_i$ to $x_i'$.
We are interested in the size distribution $f(x,t)$ of the system, related to the probability $P(x_1^{}, \dots, x_N^{}; t)$ via:
\begin{equation}
  \label{eq: size distribution}
  f(x,t) = \frac{1}{N} \int d^N\!x\, \sum_{i=1}^{N} \delta(x_i - x) P(x_1^{},\dots,x_{N}^{}; t),
\end{equation}
where $\int d^N\!x \equiv \int_0^\infty d x_{1}^{} \cdots \int_0^\infty d x_{N}^{}$. The time evolution of the distribution
$f(x,t)$ can be obtained once the transition rate is given, which also depends on whether or not new elements are produced~\cite{Choi2009,Goh2010}. Here we describe how the interactions between elements change the behavior of the distribution.

\subsection{No production}
When the total number of elements is fixed, namely, in the case of no production of new elements, one may adopt a simple process involving the size change (growth) by an amount proportional to the current size.
Accordingly, the transition rate takes the form
\begin{equation}
  \label{eq: transition rate}
  \omega(x_i \rightarrow x_i') = \lambda \delta[x_i'-(1{+}b)x_i]
\end{equation}
with the (mean) growth rate $\lambda$ and the growth factor $b$.
From Eq.~(\ref{eq: master equation}) the evolution equation for the size distribution $f(x,t)$ can be obtained straightforwardly~\cite{Goh2010}:
\begin{equation}
  \label{eq: number conserved}
  \frac{\partial{}}{\partial{t}} f(x,t) = - \lambda f(x,t) + \frac{\lambda}{1+b} f\left(\frac{x}{1+b},t \right).
\end{equation}

We now incorporate two types of interaction between elements into this transition rate. The first one to consider is that the interaction between two elements depends on the size difference between the two \cite{kim}.
This modifies the transition rate in Eq. (\ref{eq: transition rate}) and leads to
\begin{equation}
  \label{eq: transition rate A0}
  \omega(x_i \rightarrow x_i') = \lambda \delta\big[x_i'-(1{+}b)x_i - \sum_j c_{ij} (x_j {-}x_i) \big],
\end{equation}
where $c_{ij}$ measures the (size-difference) coupling strength between elements $i$ and $j$.
For simplicity, we consider globally coupled interactions, $c_{ij} = c/N$, to carry out further analytic treatment.
Defining $\alpha \equiv 1+b-c$ and the mean size $\langle x\rangle \equiv N^{-1} \sum_j x_j = \int dx x f(x,t)$, we write the transition rate depending on the mean size of elements:
\begin{equation}
  \label{eq: transition rate A}
  \omega(x_i \rightarrow x_i') = \lambda \delta \left(x_i'{-}\alpha x_i {-}c \langle x \rangle\right).
\end{equation}
Note that for positive coupling ($c>0$) the interaction tends to reduce the size of an element larger than the mean size, as manifested by the transition rate.
With Eq.~(\ref{eq: transition rate A}), it is straightforward to obtain the evolution equation for the size distribution $f(x,t)$:
\begin{align}
  \label{eq: number conserved A}
  \frac{\partial{}}{\partial{t}} f(x,t) = & - \lambda f(x,t) \theta\left(x{+}\frac{c\left<x \right>}{\alpha}\right) \nn \\
& + \frac{\lambda}{\alpha} f \left( \frac{x{-}c\left<x \right>}{\alpha},t \right) \theta(x{-}c\left<x \right>),
\end{align}
where $\theta(x)$ is the Heaviside step function.

The other type of interaction is given by the product of the sizes of two elements, which leads to the transition rate in the form
\begin{equation}
  \label{eq: transition rate B0}
  \omega(x_i \rightarrow x_i') = \lambda \delta\big[x_i'-(1{+}b)x_i + \sum_j \kappa_{ij} x_i x_j \big].
\end{equation}
Again, for simplicity, we consider the global coupling and choose the (size-product) coupling strength $\kappa_{ij} = \kappa/N$, which reduces Eq.~(\ref{eq: transition rate B0}) to
\begin{equation}
  \label{eq: transition rate B}
  \omega(x_i \rightarrow x_i') = \lambda \delta[x_i'{-}(1{+}b{-}\kappa \left<x \right>)x_i ].
\end{equation}
For positive coupling ($\kappa>0$), the size growth is suppressed by the interaction as before.
In this case, the time evolution of the distribution is described by the following equation:
\begin{align}
  \label{eq: number conserved B}
  \frac{\partial{}}{\partial{t}} f(x,t) =&\left[ - \lambda f(x,t) 
 + \frac{\lambda}{1{+}b {-}\kappa \left<x \right> } f \left( \frac{x}{1{+}b {-} \kappa \left<x \right>},t \right) \right] \nn \\
& \times \theta(1{+}b {-} \kappa \left<x \right>).
\end{align}

\subsection{Uniform size production}
When the total number of elements varies with time, i.e., $N=N(t)$, the time evolution equation needs to be modified. If each element tends to produce a new one with rate $r$ (and the total number of elements thus increases in proportion to the current number: $\dot{N} = rN$), the time evolution equation for the size distribution takes the following form~\cite{Goh2010}
\begin{equation}
  \label{eq: f evolution 2}
  \frac{\partial{f(x,t)}}{\partial{t}} = -(r+\lambda) f(x,t) + \frac{\lambda}{1+b} f \left({\frac{x}{1+b},t} \right) + rg(x,t),
\end{equation}
where $g(x,t)$ is the size distribution function of newly produced elements at time $t$. In this work we consider the case that new elements are produced in uniform size $x_0 $, i.e., $g(x,t) = \delta (x{-}1)$ after setting $x_0 \equiv 1$. 

In the presence of the interaction between elements, we may still follow the procedure similar to that in Ref. \cite{Goh2010} and derive the time evolution of the distribution function. In case that the transition rate is given by  Eq.~(\ref{eq: transition rate A}), the time evolution equation obtains the form
\begin{align}
  \label{eq: uniform}
  \frac{\partial{}}{\partial{t}} f(x,t) = &-\left[ r {+} \lambda \theta\left(x{+}\frac{c\left<x \right>}{\alpha}\right) \right] f(x,t) \nn \\
 &+ \frac{\lambda}{\alpha} f \left( \frac{x{-}c\left<x \right>}{\alpha},t \right) \theta(x{-}c\left<x \right>) + r \delta(x{-}1).
\end{align}
In the case of the transition rate given by Eq.~(\ref{eq: transition rate B}), the evolution of the distribution function is governed by
\begin{align}
  \label{eq: uniform kappa}
  \frac{\partial{}}{\partial{t}} f(x,t) = & \left[ - \lambda f(x,t) 
 + \frac{\lambda}{1{+}b {-}\kappa \left<x \right> } f \left( \frac{x}{1{+}b {-} \kappa \left<x \right>},t \right) \right] \nn \\
&\times \theta(1{+}b {-} \kappa \left<x \right>)  -r f(x,t) + r \delta(x{-}1).
\end{align}


\section{Calculation of Skewness}

To understand how the interaction affects the skew distribution emerging in the growth model,
we examine the skewness of the size distribution.
The skewness $\gamma_1$ of a distribution $f(x,t)$ is defined to be
\begin{equation}
\label{eq: def_skew}
 \gamma_1 \equiv \frac{\langle x^3 \rangle - 3\langle x \rangle \sigma_x^2 - \langle x \rangle^3}{\sigma_x^3 }\,,
\end{equation}
where $\langle X \rangle \equiv \int dx\,X f(x,t)$ is the mean (expectation) value of variable $X$ and
$\sigma_x \equiv \sqrt{\langle x^2 \rangle - \langle x \rangle^2}$ is the standard deviation.
In this work we compute $\gamma_1$, depending on $\left<x \right>$, $\left<x^2 \right>$, and $\left<x^3 \right>$, in two ways: One is direct numerical integration of the evolution equation given by Eqs.~(\ref{eq: number conserved A}), (\ref{eq: number conserved B}), (\ref{eq: uniform}), and (\ref{eq: uniform kappa}), from which we obtain the distribution function $f(x,t)$ and subsequently $\gamma_1$. In numerical integration, we use the second-order Runge-Kutta method for time integration with a time step $\delta t=0.01$, while
dividing the positive $x$ space into segments of equal length: $\delta x =1/101$.
The results obtained via numerical integration are compared with those obtained from analytic expressions of $\left< x^n \right>~(n{=}1,2,3)$ described below.
In this work we report mainly the analytic results unless specified otherwise, since the two methods yield essentially the same results. 

\subsection{Size-difference interaction}
We first consider the interaction depending on the difference between sizes, which gives
the transition rate in Eq.~(\ref{eq: transition rate A}), and examine the evolution of the system in the presence of uniform production as well as in the absence of production. 
To compute the skewness $\gamma_1$, we construct the time evolution equation for $\langle x^n \rangle~(n=1,2,3)$.
Taking the time derivative 
\begin{equation}
\label{eq: diff mean x u}
  \frac{d\langle x \rangle}{dt}  = \int dx\, x \frac{\partial }{\partial t}f(x,t)
\end{equation}
and making use of Eq.~(\ref{eq: uniform}), we obtain
\begin{equation}
  \frac{d\left<x\right>}{dt}  = u \left<x\right> + r
\end{equation}
with $u\equiv\lambda b- r$.
Note that the coupling strength $c$ is not involved in determining $\left<x \right >$. With the initial value $\left<x \right >_0$, the solution is found to be 
\begin{equation}
\label{eq: mean x u}
\left<x \right> = A_u e^{u t}+A_c\,,
\end{equation}
where $A_u \equiv \left<x \right >_0-A_c$ and $A_c \equiv -r/u$.
Similarly, we obtain the differential equation for $\langle x^2 \rangle$:
\begin{equation}
\label{eq: dxsq u}
  \frac{d\langle x^2 \rangle}{dt} = v \langle x^2 \rangle + J \left<x \right >^2  +r \,,
\end{equation}
where $v \equiv \lambda (\alpha^2 -1)-r$ and $J \equiv \lambda c (2 \alpha +c)$.
Putting Eq.~(\ref{eq: mean x u}) into Eq.~(\ref{eq: dxsq u}) leads to the solution
\begin{equation}
\label{eq: xsq u}
  \langle x^2 \rangle = B_v e^{v t} +  B_1 e^{u t}+B_2 e^{2u t}+B_c\,,
\end{equation}
where $B_v \equiv \langle x^2 \rangle_{0}-B_1 -B_2 -B_c$ with
\begin{equation}
B_1 \equiv \frac{2JA_c A_u}{u-v},~ B_2 \equiv \frac{J A_u^2}{2u-v},~\mbox{and}~ B_c \equiv  \frac{J A_c^2 +r}{-v}.
\nn
\end{equation}

Again in a similar manner, we obtain the expression for $d\langle x^3 \rangle/dt$. Denoting
$w \equiv \lambda (\alpha^3-1)-r$, $z \equiv u+v$, $K \equiv 3 \lambda c \alpha^2$, and $M \equiv \lambda c^2 (3 \alpha+c)$, we have
\begin{equation}
\label{eq: dxcube u}
  \frac{d}{dt}\langle x^3 \rangle = w \langle x^3 \rangle +K\langle x^2 \rangle\left<x \right>+ M\left<x \right>^3 + r \,,
\end{equation}
the solution of which reads
\begin{align}
\label{eq: xcube u}
  \left<x^3 \right > & =  C_w e^{wt}+C_z e^{zt}+C_v e^{v t} \nn \\
 & \quad +  C_1 e^{u t}+C_2 e^{2u t}+C_3 e^{3u t}+C_c
\end{align}
with
\begin{align}
   C_1 &\equiv \frac{K(A_c B_1 + A_u B_c)+3MA_c^2A_u}{u-w}, \nn \\
   C_2 &\equiv \frac{K(A_c B_2 + A_u B_1)+3MA_c A_u^2}{2u-w}, \nn \\
   C_3 &\equiv\frac{K A_u B_2 +MA_u^3}{3u-w}~,  C_c =\frac{K A_c B_c+MA_c^3+r}{-w}\nn \\
   C_z &\equiv\frac{K A_u B_v}{z-w},~ ~ C_v = \frac{K A_c B_v}{v-w}, \nn
\end{align}
and $C_w \equiv \langle x^3 \rangle_{0} -C_z-C_v -C_1 -C_2 -C_3-C_c$.

In the absence of production, we have $r= A_c=0$, which in turn lead to $B_1 = B_c = C_v =C_ 2=C_c =0$ in the expressions of $\langle x^2 \rangle$ and $\langle x^3 \rangle$. Especially, when there is no interaction or $c=0$, setting $\left<x \right >_{0}=1$, Eqs. (\ref{eq: mean x u}), (\ref{eq: xsq u}), and (\ref{eq: xcube u}) are further simplified as

\begin{align}
\label{mean x A}
 &\left<x \right > =  e^{\lambda b t}, \nn \\
 &\left<x^2 \right > =  \left<x^2 \right >_{0}  e^{\lambda b(b+2) t},  \\
 &\left<x^3 \right > =   \left<x^3 \right >_{0} e^{\lambda b(b^2+3b+3) t}. \nn \\
\nn
\end{align}

Now using Eq. (\ref{eq: def_skew}) with the above expressions for
$\left<x \right >$, $\langle x^2 \rangle$ and $\langle x^3 \rangle$, one can calculate skewness of the size distribution without \textit{any} adjustable parameters.
When we compare the results of the analytic expressions with those obtained via numerical integration, we use the same initial values $\langle x^2 \rangle_{0}$ and $\langle x^3 \rangle_{0}$.

\subsection{Size-product interaction}
We next consider the interaction depending on the product of sizes, which corresponds to the transition rate given by Eq.~(\ref{eq: transition rate B}), and compute
the skewness of the size distribution in a similar way.
In case that no elements are produced newly, the evolution equations for $\langle x^n \rangle$ read $(n=1,2,3)$:
\begin{equation}
\label{eq: diff mean xi kap no}
  \frac{d}{dt} \langle x^n \rangle = \lambda [(1+b-\kappa \left<x \right >)^n -1] \langle x^n \rangle.
\end{equation}
For $n=1$, Eq.~(\ref{eq: diff mean xi kap no}) bears the solution
\begin{equation}
\label{eq: mean x kap no}
  \left<x \right > =  \frac{b e^{\lambda b t}}{\psi}\,,
\end{equation}
where $\psi \equiv b+\kappa (e^{\lambda bt}-1)$ depends on $t$.
Inserting this into Eq.~(\ref{eq: diff mean xi kap no}) for $n=2$ and $3$, we obtain
\begin{align}
\label{eq: mean x2 kap no}
 \langle x^2 \rangle =&\langle x^2 \rangle_0 e^{(b+2)\log b + \kappa -b} \nn \\
      &\times \exp\left[\frac{b(b-\kappa)}{\psi}+\lambda b(b+2)t -(b+2) \log\psi \right] \\
\label{eq: mean x3 kap no}
\langle x^3 \rangle =&\langle x^3 \rangle_0 e^{(b^2 +3b+3)\log b -(3b^2 /2)-3b+2b\kappa + 3\kappa-(\kappa^2 /2)}\nn \\
 &\times \exp \Big[\frac{b(b^2 +b(3-\kappa)-3\kappa)}{\psi}+ \frac{b^2 (b-\kappa)^2}{2\psi^2} \nn\\
 &~~~~~~~~ +(b^2+3b+3)(\lambda bt - \log\psi )\Big].
\end{align}

When new elements of uniform size are produced,
the evolution equation for $\langle x^n \rangle$ takes the form $(n=1,2,3)$
\begin{equation}
\label{eq: diff mean xi kap u}
 \frac{d}{dt} \langle x^n \rangle = -(r+\lambda) \langle x^n \rangle +\lambda(1+b-\kappa \left<x \right>)^n \langle x^n \rangle + r ,
\end{equation}
which is solvable for $n=1$ to give the analytic solution for $\left<x \right>$.
When $\kappa>0$, it is given by
\begin{equation}
\label{eq: mean x kap p u}
  \left<x \right > =  \frac{u}{2 \chi  } +\frac{S}{2 \chi} \tanh \left[\frac{S}{2  }t -\tanh^{-1} \frac{2 \chi -u}{S} \right]
\end{equation}
with $u\equiv\lambda b - r$, $\chi \equiv \kappa \lambda$, and $S\equiv\sqrt{u^2 + 4 \chi r}$.
When $\kappa <0$, the solution reads
\begin{equation}
\label{eq: mean x kap n u}
  \left<x \right > =  \frac{S-u + (S+u)C e^{St}}{2 \chi (C e^{S t}-1) }
\end{equation}
with $C\equiv(u-2 \chi-S)/(u-2 \chi+S)$. To find higher moments $\langle x^n\rangle$ for $n=2$ and $3$, one should resort to numerical methods. 
We thus put Eq.~(\ref{eq: mean x kap p u}) or (\ref{eq: mean x kap n u}) into
Eq.~(\ref{eq: diff mean xi kap u}) and perform numerical integration.

\section{Results and Discussions}

\subsection{No production}
\begin{figure}
\includegraphics[width=8cm]{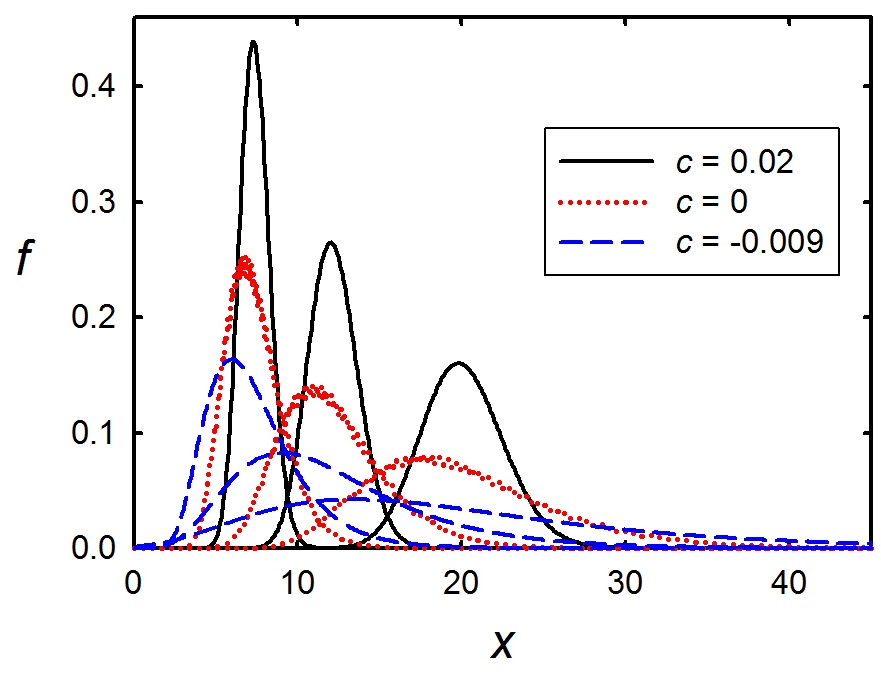}
\caption{(Color online) Evolution of the size distribution $f(x)$ in the case of no production ($r=0$) for $\lambda = 0.2$ and $b=0.025$. Data have been obtained via numerical integration of Eq.~(\ref{eq: number conserved A}) for three values of the size-difference coupling $c$, as shown in the legend.
Three successive distributions are shown at time $t=400$, $500$, and $600$ for each value of $c$.}
\label{fig: fvsx_c_no}
\end{figure}

When no new elements are produced ($r=0$), the size distribution $f(x)$, obtained via numerical integration of Eq.~(\ref{eq: number conserved A}) for $\lambda = 0.2$ and $b=0.025$, is shown in Fig.~\ref{fig: fvsx_c_no}
for three values of the coupling: $c = 0.02, 0$, and $-0.009$.
Three successive distributions are plotted at time $t=400$, $500$, and $600$ for each value of $c$, illustrating the time evolution. It is observed that at given time, as the interaction of positive/negative coupling is turned on, the peak heights and positions increase/decrease and move towards larger/smaller sizes, respectively, with the skewness becoming smaller/larger.
As time goes on, the peak height and position become eventually decreasing and moving towards larger size, regardless of the interaction. On the other hand, the skewness tends to increase for negative coupling while it appears to change little for positive coupling during the time shown in the figure.

\begin{figure}
\includegraphics[width=7.5cm]{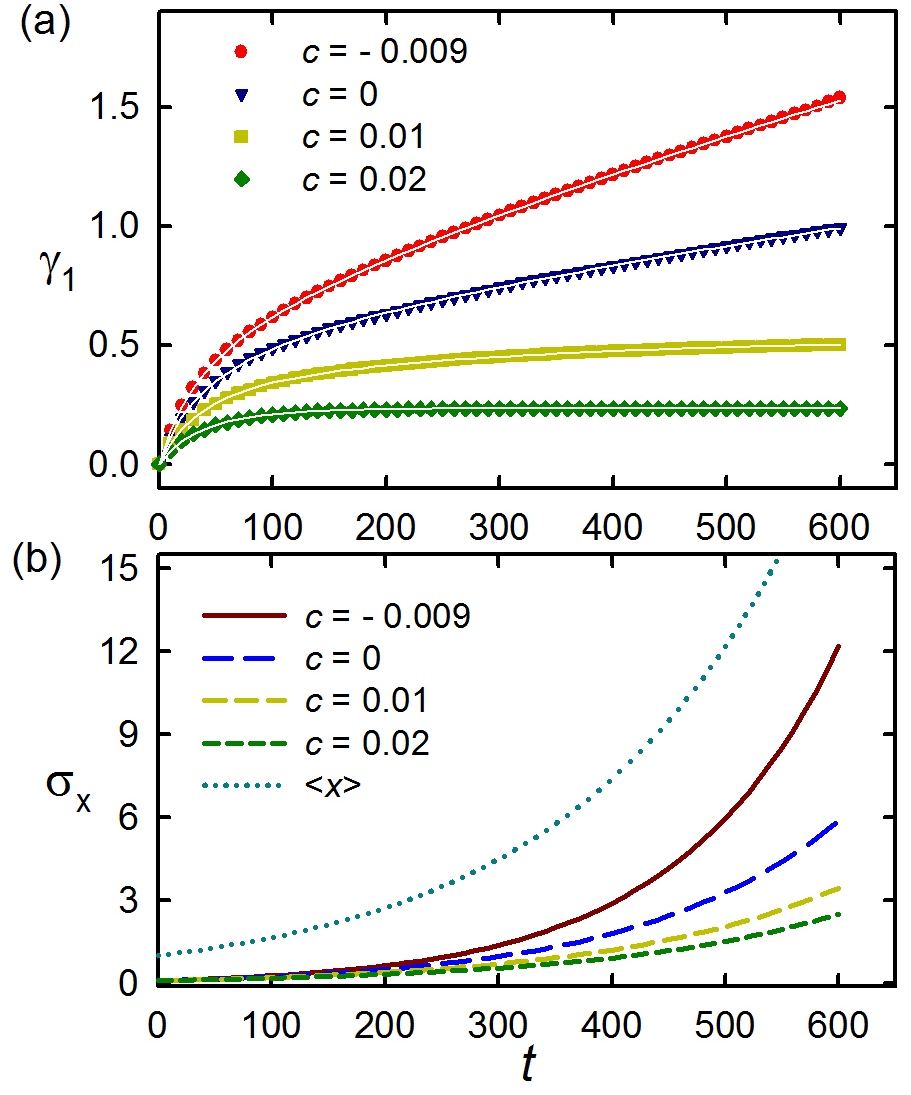}
\caption{(Color online) (a) Skewness $\gamma_1$ and (b) standard deviation $\sigma_x$ of the size distribution versus time $t$ for four values of coupling $c$ shown in the legend.
Symbols represent data points obtained via numerical integration for the same system as in Fig.~\ref{fig: fvsx_c_no}.
Lines and white lines passing through symbols in (a) depict the analytical results from Eqs. (\ref{eq: def_skew}) and (\ref{mean x A}).
Also plotted in (b) is the mean size $\left<x\right>$ as a dotted line.}
\label{fig: skvstc}
\end{figure}

This is elucidated more in Fig.~\ref{fig: skvstc}, which displays the mean size, standard deviation, and skewness of the distribution. In Fig.~\ref{fig: skvstc}(a), we show the evolution of
the skewness $\gamma_1$ for four different values of coupling $c$ shown in the legend.
Symbols designate data points obtained via numerical integration for the same system as in Fig.~\ref{fig: fvsx_c_no}; lines and white lines passing through symbols in (a) denote the analytical results from the analytic expressions derived in the previous section.
Plotted in Fig.~\ref{fig: skvstc}(b) is the standard deviation $\sigma_x$ versus time $t$
as well as  $\left<x\right>$ (dotted line), which is independent of $c$.
In general the skewness increases with time but the increase is slower for larger values of $c$.
At the largest value $c=0.02$, in particular, the skewness appears to saturate after the initial increase.

\begin{figure}
\includegraphics[width=8cm]{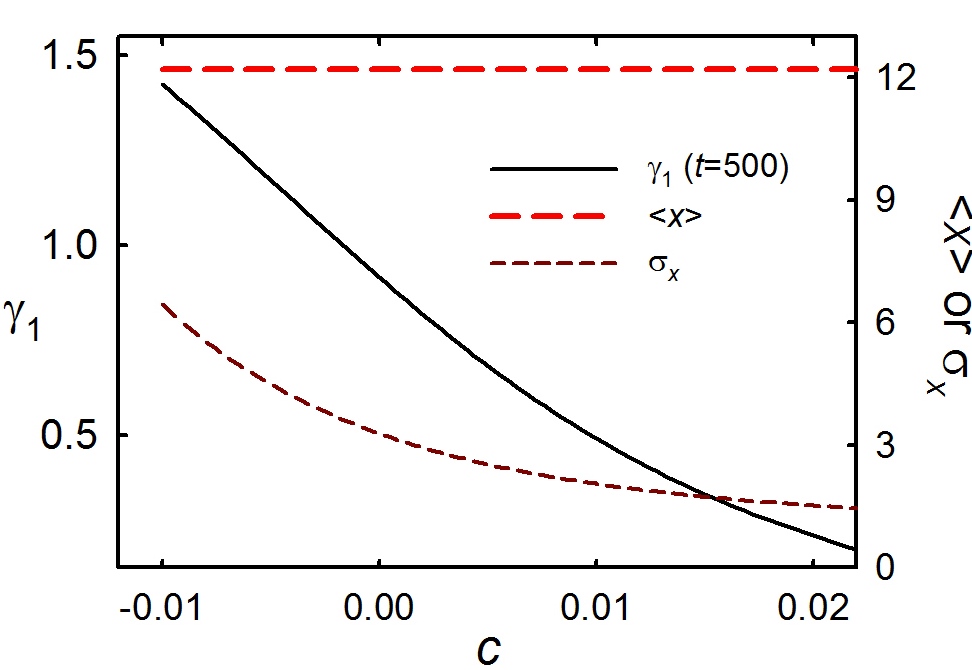}
\caption{(Color online) Skewness $\gamma_1$, mean size $\left<x\right>$, and standard deviation $\sigma_x$ versus coupling $c$ at time $t=500$, obtained from the analytical expressions for the same system as in Fig.~\ref{fig: fvsx_c_no}.
}
\label{fig: vsc_no}
\end{figure}

Figure~\ref{fig: vsc_no} exhibits the skewness $\gamma_1$, mean size $\left<x\right>$, and standard deviation $\sigma_x $ versus coupling strength $c$ at time $t=500$, obtained from analytic expressions.
It is observed that the skewness tends to decrease as the coupling strength is increased.
This behavior is understood if we examine the transition rate in Eq.~(\ref{eq: transition rate A}):
In a noninteracting system, the skewness as well as the mean size increases with time, as manifested in Fig.~\ref{fig: fvsx_c_no}. The interaction of positive coupling ($c>0$) suppresses the size growth, resulting in the decrease of skewness. This may be inferred from the transition rate in Eq.~(\ref{eq: transition rate A0}), where the mean value of the total interaction $\sum_j c_{ij} (x_j -x_i )$ vanishes for constant coupling $c_{ij}=c/N$, thus establishing the independence of $\left< x \right>$ on $c$.

\begin{figure}
\includegraphics[width=8cm]{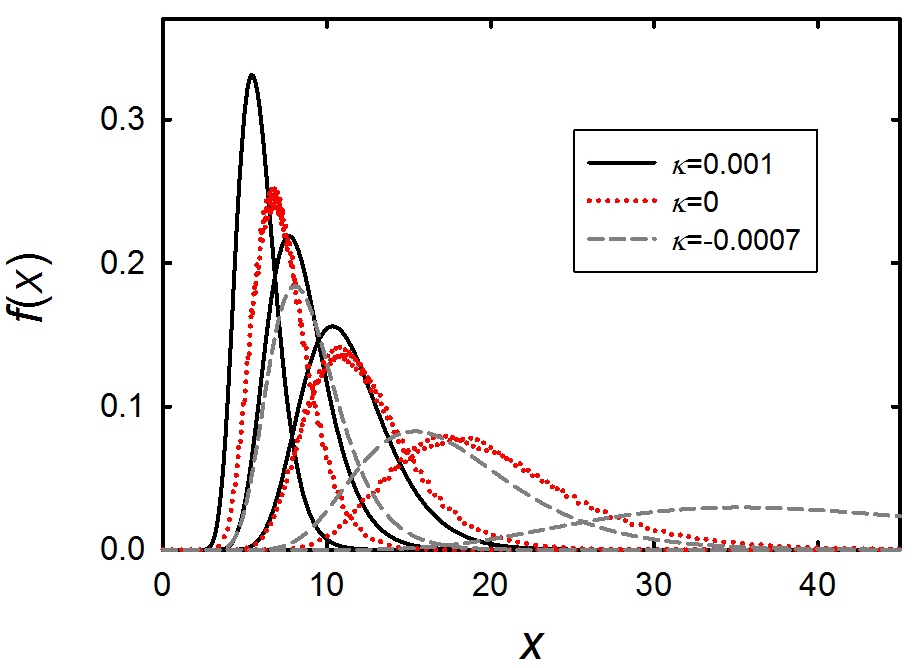}
\caption{(Color online)
Evolution of the size distribution $f(x)$ in the case of no production ($r=0$) for $\lambda = 0.2$ and $b=0.025$. Data have been obtained via numerical integration of Eq.~(\ref{eq: number conserved B}) for three values of the size-product coupling $\kappa$, as shown in the legend.
Three successive distributions are shown at time $t=400$, $500$, and $600$ for each value of $\kappa$.}
\label{fig: fvsx_K_no}
\end{figure}

In the case of the transition rate given by Eq.~(\ref{eq: transition rate B}), the interaction effects are similar to what have been described above.
Figure~\ref{fig: fvsx_K_no} displays the size distribution $f(x)$ for $\lambda = 0.2$ and $b=0.025$,
obtained via numerical integration of Eq.~(\ref{eq: number conserved B}) for three values of coupling $\kappa$ shown in the legend. Again three successive distributions are shown at time $t=400$, $500$, and $600$ for each value of $\kappa$.
It is observed that, as $\kappa$ is increased/decreased from zero, the peak height in general increases/decreases with the skewness being decreased/increased at each given time.
However, the peak position moves toward smaller/larger values of $x$, in contrast to the case of Eq.~(\ref{eq: transition rate B}).
Is is also observed that, as time proceeds, the peak height decreases and the peak position moves towards larger sizes, regardless of the interaction. The skewness appears to increase monotonically for all values of coupling $\kappa$, but the rate of increase becomes, for positive $\kappa$, smaller at large times.

\begin{figure}
\includegraphics[width=7.5cm]{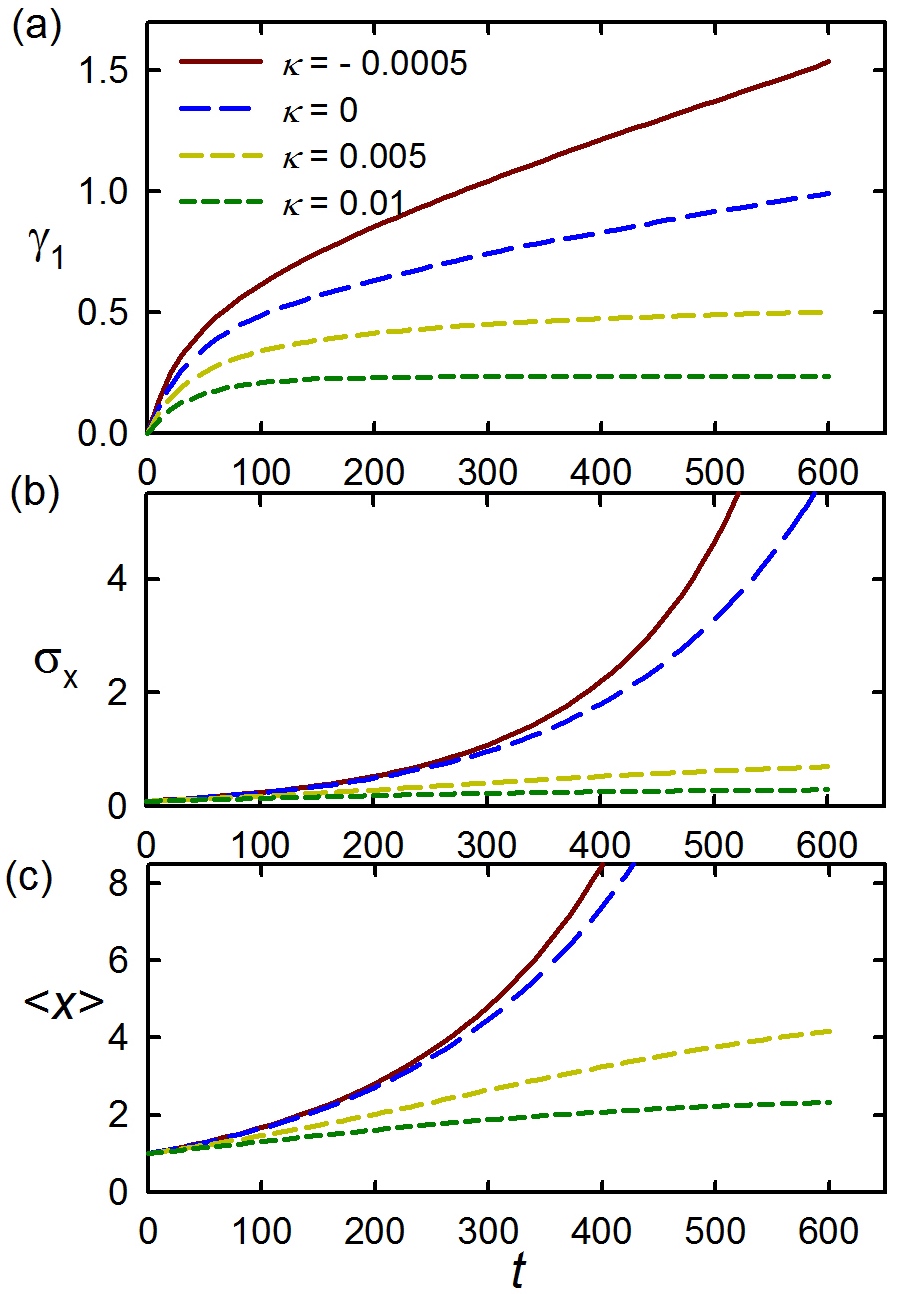}
\caption{(Color online) Analytical results of (a) skewness $\gamma_1$, (b) standard deviation $\sigma_x$, and (c) mean $\left<x\right>$ versus time $t$ for four values of coupling $\kappa$ shown in the legend.
Other parameters are the same as those in Fig.~\ref{fig: fvsx_K_no}.
}
\label{fig: skvstK}
\end{figure}

We look more closely at this behavior in Fig.~\ref{fig: skvstK}, which displays
(a) skewness, (b) standard deviation, and (c) mean of the distribution versus
time $t$ for four values of coupling $\kappa$, obtained from Eqs.~(\ref{eq: mean x kap no})-(\ref{eq: mean x3 kap no}).
These quantities grow with time for all values of $\kappa$ except for the apparent saturation for $\kappa>0$ at large time.
This is to be contrasted with the case of the size-difference interaction, in which only the skewness $\gamma_1$ tends to saturate for positive coupling ($c>0$) (see Fig.~\ref{fig: skvstc}).
Further at given time, the mean size does depend on $\kappa$ and becomes saturated for positive values while it does not depend on the value of $c$ (not shown).

\subsection{Uniform size production}

\begin{figure}
\includegraphics[width=8cm]{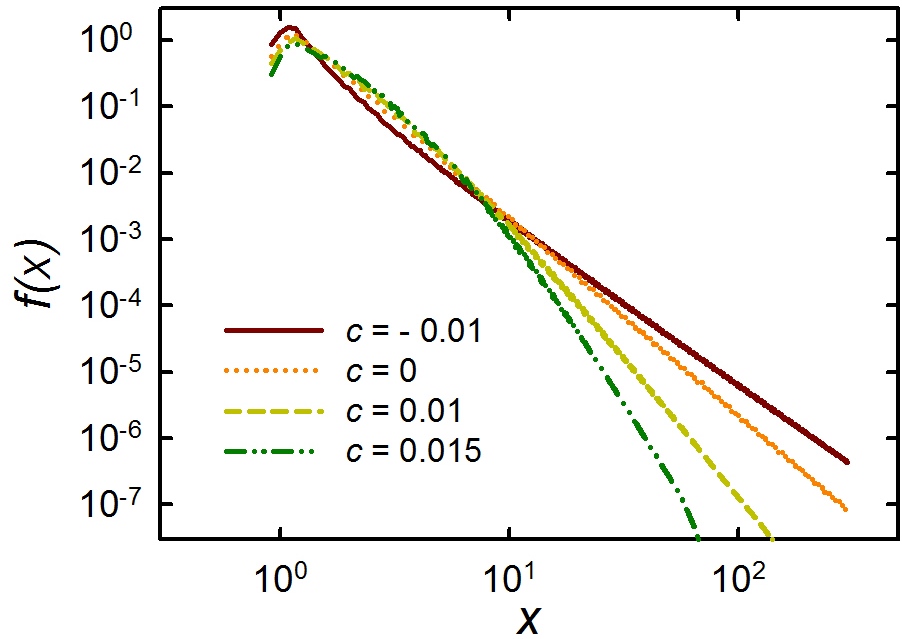}
\caption{(Color online) Size distribution $f(x)$ in the case of uniform production for $\lambda = 0.4$, $b=0.025$ and $r=0.02$ at time $t=700$. Data have been obtained via numerical integration of Eq.~(\ref{eq: uniform}) for four values of size-difference coupling $c$, as shown in the legend.
 }
\label{fig: fvsx_u}
\end{figure}

In this subsection, we discuss the effects of interaction between elements in case that new elements of uniform size $x_0\, (\equiv 1)$ are produced, i.e., $g(x,t)=\delta (x{-}1)$. In the absence of interaction, the size distribution obeys the power law in the long-time limit \cite{Goh2010}:
\begin{equation}
 \label{eq: stationary power}
 f(x) \sim x^{-\eta}
\end{equation}
with the exponent
\begin{equation}
\label{eq: exponent}
\eta = 1 + \frac{\ln{(1+r/\lambda)}}{\ln{(1+b)}}. \nn
\end{equation}

\begin{figure}
\includegraphics[width=7.5cm]{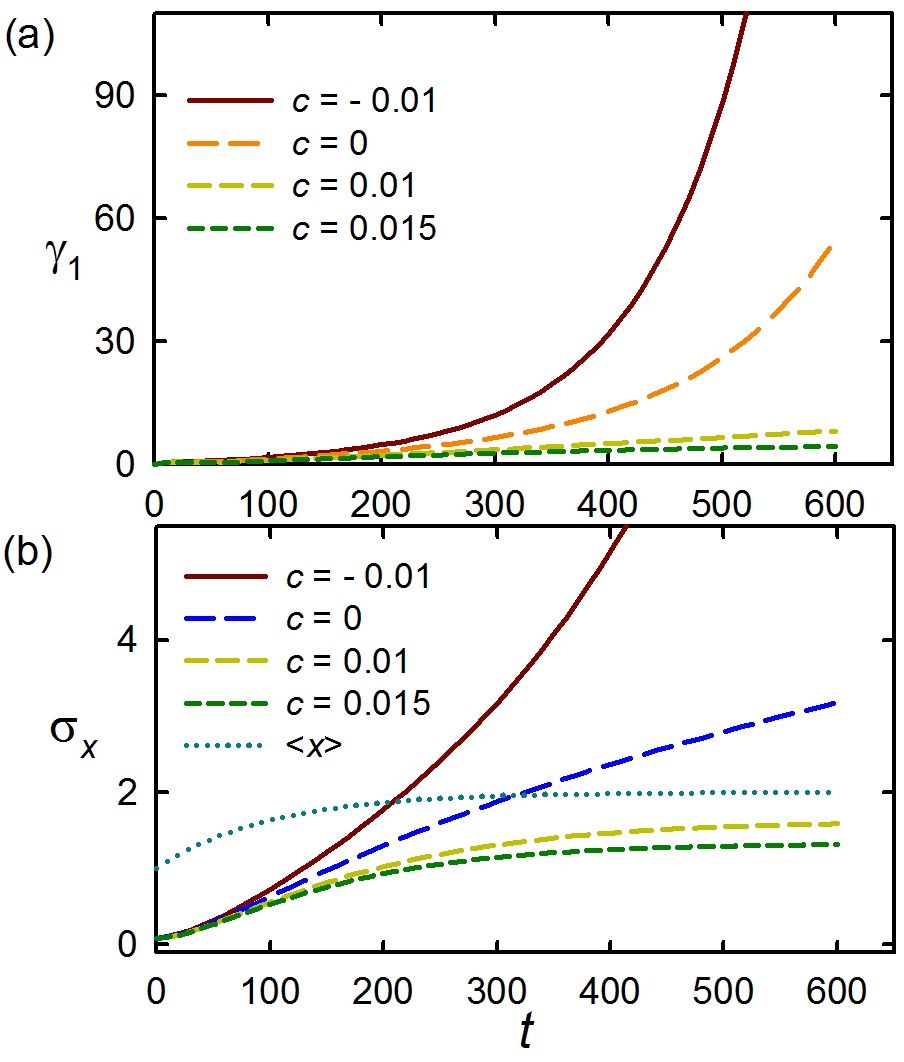}
\caption{(Color online) Analytical results for the same system as in Fig.~\ref{fig: fvsx_u}: (a) skewness $\gamma_1$ and (b) standard deviation $\sigma_x$ versus time $t$ for four values of coupling $c$ shown in the legend.
Also plotted in (b) is the mean size $\left<x\right>$ as a dotted line.}
\label{fig: skvstcu}
\end{figure}

To probe how interactions alter such power-law behavior, we carry out numerical integration of Eq.~(\ref{eq: uniform}) with $\lambda = 0.4$, $b=0.025$, and $r=0.02$, and display in Fig.~\ref{fig: fvsx_u} the resulting size distribution $f(x)$ at time $t=700$ 
for four values of size-difference coupling $c$, as shown in the legend.
As revealed in Fig.~\ref{fig: fvsx_u}, the presence of interaction induces the distribution $f(x)$ to deviate from the power law; the way how it deviates depends on the sign of $c$.
Note that the power-law distribution in Eq.~(\ref{eq: stationary power}) with the exponent $\eta \leq 3$ possesses diverging skewness. With this in mind, we investigate the interaction effects on the skewness of the distribution.

\begin{figure}
\includegraphics[width=8.5cm]{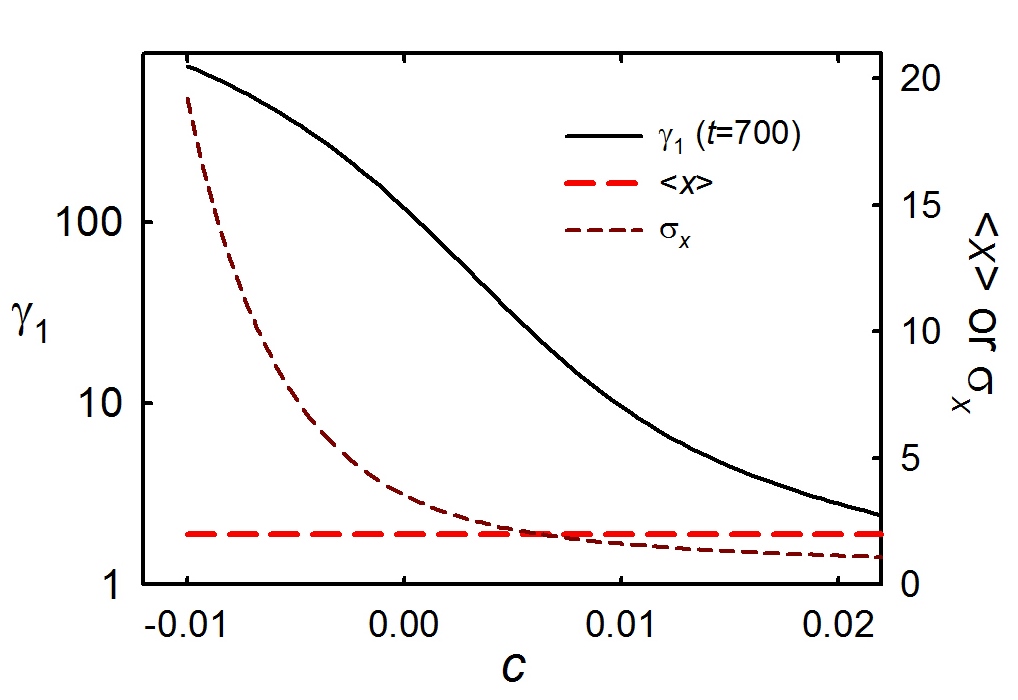}
\caption{(Color online) Skewness $\gamma_1$, mean size $\left<x\right>$, and standard deviation $\sigma_x$ versus coupling $c$ at time $t=700$,
obtained from analytic expressions for the same system as in Fig.~\ref{fig: fvsx_u}.
}
\label{fig: vsc_u}
\end{figure}

Figure~\ref{fig: skvstcu} shows the time evolution of (a) skewness $\gamma_1$ and (b) standard deviation $\sigma_x$ for four values of coupling $c$ as shown in the legend,
which have been obtained from analytic expressions for the same system as in Fig.~\ref{fig: fvsx_u}.
While the skewness $\gamma_1$ diverges for $c\leq 0$, it increases slowly and appears to approach a finite value for $c>0$. This behavior of skewness changing with $c$
is consistent with the gradients of the distribution curves in Fig.~\ref{fig: fvsx_u}.
In this system, the mean size $\left<x\right>$, plotted with a dotted line in Fig. \ref{fig: skvstcu}(b), approaches a steady-state value, independently of the coupling $c$.
These features are illustrated in Fig.~\ref{fig: vsc_u}, which presents analytical results of skewness $\gamma_1$, mean size $\left<x\right>$, and standard deviation $\sigma_x$ versus $c$ at time $t=700$.

\begin{figure}
\includegraphics[width=7.5cm]{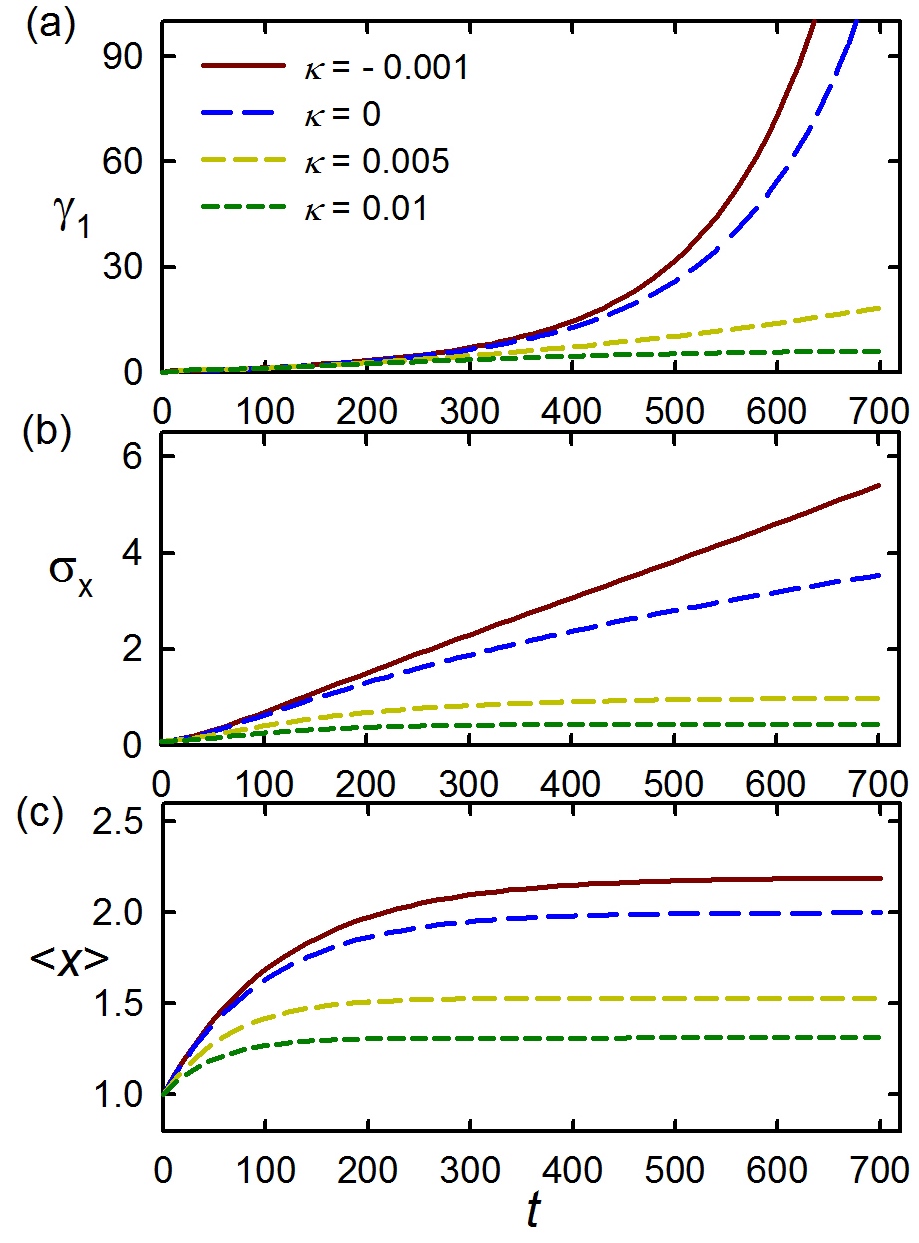}
\caption{(Color online) (a) Skewness $\gamma_1$, (b) standard deviation $\sigma_x$ and (c) mean size $\langle x\rangle$ of the distribution versus time $t$ in a system with uniform size production for $\lambda = 0.4$, $b=0.025$, $r=0.02$, and four values of size-product coupling $\kappa$ as shown in the legend. These have been calculated by means of a semi-analytic method, described in the text.
}
\label{fig: skvstK-u}
\end{figure}

Finally, we briefly describe the effects of the other type of interaction depending on the size product, which corresponds to Eq.~(\ref{eq: transition rate B}).
Figure~\ref{fig: skvstK-u} exhibits the time evolution of (a) skewness $\gamma_1$, (b) standard deviation $\sigma_x$, and (c) mean size $\langle x\rangle$ of the distribution in the presence of uniform size production, for $\lambda = 0.4$, $b=0.025$, $r=0.02$, and
four values of size-product coupling $\kappa$ shown in the legend.
These have been calculated semi-analytically through the use of Eqs.~(\ref{eq: diff mean xi kap u}), (\ref{eq: mean x kap p u}), and (\ref{eq: mean x kap n u}). The overall features are more or less the same: Skewness $\gamma_1$ decreases as coupling $\kappa$ is increased. The standard deviation as well as the skewness grows with time and becomes saturated at large time for $\kappa>0$.
The mean size also grows with time and becomes saturated even for negative values of $\kappa$ considered here. The mean size saturated at given time tends to decrease as $\kappa$ is increased.

\section{Summary}

We have studied the effects of interaction between elements in a growth model, governed by the master equation. Specifically, we have incorporated two types of interaction into the transition rate: one type depending on the difference between element sizes and the other depending on the product of sizes. In the global-coupling limit, we have developed the evolution equation for the size distribution function in terms of the growth rate, growth factor, coupling constant between elements, and the production rate of new elements. From the evolution equations for the size we have derived analytic expressions for the first three moments of size variables, $\left<x \right>$, $\langle x^2 \rangle$, and $\langle x^3 \rangle$, from which the standard deviation and the skewness have been computed.
In the case of no production, for which the asymptotic distribution is the log-normal distribution, it has been found that the mean size, standard deviation, and skewness decrease as the coupling constant $c$ or $\kappa$ is varied from negative to positive values, with apparent saturation of the mean size at large time. In the case of production of new elements, for which the asymptotic distribution is the power-law distribution, deviations from the power law with the decrease of the mean size, standard deviation, and skewness have been observed as the coupling $c$ or $\kappa$ is varied from negative to positive values. Notably the mean size of the distribution saturates at large time for all vales of $c$ and $\kappa$ considered.

\section*{ACKNOWLEDGMENT}
This work was supported in part by the 2016 Research Fund of the University of Ulsan.

\end{document}